\documentstyle[12pt]{laa}

\def\aa #1 #2 {A\&A {#1} #2}
\def\aas #1 #2 {A\&AS {#1} #2}
\def\araa #1 #2 {ARA\&A {#1} #2}
\def\mon #1 #2 {MNRAS {#1} #2}
\def\apj #1 #2 {ApJ {#1} #2}
\def\apjs #1 #2 {ApJS {#1} #2}
\def\apjl #1 #2 {ApJ {#1} #2}
\def\aj #1 #2 {AJ {#1} #2}
\def\ltsima{$\; \buildrel < \over \sim \;$}
\def\lsim{\lower.5ex\hbox{\ltsima}}
\def\gtsima{$\; \buildrel > \over \sim \;$}
\def\gsim{\lower.5ex\hbox{\gtsima}}

\def\aa #1 #2 {A\&A, {#1}, #2}
\def\aas #1 #2 {A\&AS, {#1}, #2}
\def\araa #1 #2 {ARA\&A, {#1}, #2}
\def\mon #1 #2 {MNRAS, {#1}, #2}
\def\apj #1 #2 {ApJ, {#1}, #2}
\def\apjs #1 #2 {ApJS, {#1}, #2}
\def\apjl #1 #2 {ApJ, {#1}, #2}
\def\aj #1 #2 {AJ, {#1}, #2}
\def\nat #1 #2 {Nature, {#1}, #2}
\def\pasj #1 #2 {PASJ, {#1}, #2}
\def\pasp #1 #2 {PASP, {#1}, #2}

\input psfig.sty

\begin{document}


\title{SIGMA observations of X-ray Nova Velorum 1993 (GRS 1009$-$45)}

 \author{P. \ts Goldoni\inst{1}, M. \ts Vargas\inst{1},
         A. \ts Goldwurm\inst{1}, P. Laurent\inst{1},
         J.-P. \ts Roques\inst{2}, E. \ts Jourdain\inst{2},
         J.\ts Malzac\inst{2}, G. \ts Vedrenne\inst{2}, 
         M.\ts Revnivtsev\inst{3}, E.\ts Churazov\inst{3,4},
         M.\ts Gilfanov\inst{3,4}, R.\ts Sunyaev\inst{3,4},
         A.\ts Dyachkov\inst{3}, N. Khavenson\inst{3},
         I.\ts Tserenin\inst{3}, N. \ts Kuleshova\inst{3}
}

\institute{$^1$ CEA/DSM/DAPNIA, SAp, CEA-Saclay, F-91191 Gif-sur-Yvette, France; e-mail: (paolo,mv,ag,fil)@mir.saclay.cea.fr\\
$^2$ 
Centre d'Etude Spatiale des Rayonnements, 9 Avenue du Colonel Roche, BP 4346, 31029 Toulouse Cedex France\\
$^3$ 
Space Research Institute, Profsouznaya 84/32, Moscow 117810, Russia\\
$^4$ 
Max-Planck-Institut f\"ur Astrophysik, Karl-Schwarzschild-Str. 1,85740 Garching
bei Munchen, Germany\\
}

\date{Received 17/7/1997, Accepted 22/8/1997}

\maketitle

\begin{abstract}

We report on hard X--ray observations of X-ray Nova Velorum
1993 (GRS 1009-45) performed with the SIGMA coded mask X--ray telescope
in January 1994. The source was clearly detected with a flux of about
60 mCrab in the 40-150 keV energy band during the two observations with a
hard spectrum ($\alpha \sim - 1.9$) extending up to $\sim$ 150 keV.
These
observations confirm the duration of the activity of the source in hard X--rays over 100 days after the first maximum and suggest a spectral hardening which has already been observed in Nova Muscae.
These and other characteristics found in these observations strengthen the case
for this Nova to be a black hole candidate similar to Nova Muscae.

\end{abstract}

\section {Introduction}
X--ray Novae have provided some of the best galactic black hole candidates
from their dynamical mass determinations. Among them Nova Velorum 1993
(GRS 1009-45) is one of the least known. It was detected by GRANAT/WATCH
between the 12th and the 13th September 1993 (Lapshov et al., 1993) and
subsequently confirmed by BATSE (Fishman et al., 1989) with an improved
source location (Harmon et al., 1993). It appeared first in the 20-60 keV
band at 0.18 Crab reaching shortly after 0.7-0.8 Crab in the 8-60 keV total
band of the WATCH instrument.


\noindent BATSE detected the source at the same level between 20 and 500 keV
and the spectrum was a power law with a photon index $\alpha \sim -2.5$.

\noindent About a month later, in October,
the TTM team established that the source spectrum
was similar to the ones of GS2000+25 and Nova Muscae, both dynamically proven
black hole candidates (Kaniowsky et al., 1993). In the 2-200 keV band, the object presented a soft component approximated with a black body spectrum with kT$\sim $0.52 $\pm 0.03$ keV which dominates between 2 and 10 keV. At higher energies the spectrum
is well approximated by a power law with a photon index $\alpha \sim -
2.53 \pm 0.05$ (Sunyaev et al., 1994). The source had a 1 Crab flux at 3
keV and a 0.1 Crab flux at 20 keV indicating clearly the softness of its
emission (Kaniowsky et al., 1993).

\noindent In November 
the flux decreased by 25 $\% $
in the 2-6 keV band
(at 0.5 Crab) but it remained soft (Borozdin et al., 1993). No reliable
absorption column density could be established. However the column density can be considered less than about 10$^{23}$ cm$^{-2}$ due to the absence of
heavy absorption features in the spectrum (K. Borozdin, private communication).

\noindent A week later ASCA confirmed this result giving a blackbody spectrum
with kT $\sim$ 0.6 keV and F$_x \sim$ 0.8 Crab (1-10 keV) (Tanaka Y., 1993).
The absorption column density found in this observation is 3 (-0.14,+0.22)
$\times $ 10$^{21}$ cm$^{-2}$ (Ueda Y., private communication).
This value points toward a relatively nearby source with a distance
of a few kiloparsec.

\noindent The optical identification was performed by Della Valle and
Benetti (1993): they identified the source with a V$\sim $14.6 late-G/early K
type star with a broad double-peaked H$\alpha$ emission on 17 Nov. 1993 at the
ESO/MPI telescope. Baylin \& Orosz (1995) observed the source between 
150 and 200 days after the outburst and found evidence for a
secondary optical outburst and for repeated mini outbursts,
similar to those observed in Nova Persei 1992, with V-mag varying between 
16 and 20. Very recently Shahbaz et al. (1996) have reported the
discovery of an orbital, ellipsoidal modulation with an amplitude of 0.23
magnitudes and a period of 6.86 $\pm$ 0.12 hrs.
However it is not possible to derive a mass function yet as there is
no available radial velocity curve for the secondary. 

\noindent High energy ($>$20 keV) monitoring of the source have been
performed by BATSE which produced a light curve and spectral indices
for the source in the 20-100 keV band during about 150 days following the 
first maximum (Paciesas et al., 1995, Harmon et al., 1994). The overall form
of the light curve is similar to the Nova Muscae 1991 one, with two later
maxima in mid-October ($\sim$ 30 days after the primary) and in early December 
($\sim$ 85 days after the primary), although the Nova Muscae maxima occurred
later (Paciesas et al., 1995). The spectrum softens during the primary rise
reaching $\alpha \sim - $2.5 and then remains relatively constant within
errors until the December maximum. During the latter the spectrum 
becomes harder (although the statistics is worse)
reaching $\alpha \sim -$2.1 and remains at that level
until the end of observations in January (Paciesas et al., 1995, see their 
Figure 3).

\section {Observations, Data Analysis and Results}

\noindent The French coded mask telescope SIGMA provides high resolution
images n the hard-X/soft $\gamma$-ray band from 30 keV to 1300 keV, with a
typical angular resolution of 15' and a 20 hour exposure sensitivity of
$\sim$ 26 mCrab in the 40-150 keV band (Paul et al., 1991).
Launched the 1st December 
1989 onboard the Russian Granat Space observatory, SIGMA detected a 
fair number of galactic X--ray novae during its seven year lifetime
(Roques et al., 1994, Gil'fanov et al., 1993a, Trudolyubov et al., 1996,
Vargas et al., 1996, Vargas et al, 1997, Revnivtsev et al., 1997). This
class of sources is one of the primary targets of this telescope due
to its high precision localization capability unprecedented in this
energy band.

\noindent The observations we report were performed after the third outburst,
about 120 days after its discovery, on the 11th and 12th of January 1994
(JD 2449364-5).
The campaign consisted of two distinct observation sessions with the
telescope pointed in the source direction, both performed in
spectral-imaging mode (Paul et al., 1991), for a total effective time
of $\sim$ 35 hours. The parameters of the observation are listed in Table 1.


\begin{table}

\caption{SIGMA observation log of GRS 1009-45}
\label{Table1}
\[
    \begin{array}{ccc}
    \hline
\noalign{\smallskip}
  ${\rm Session}$ & ${\rm Date(U.T.)} $ & ${\rm Exposure~(hrs)}$ \\
\noalign{\smallskip}
\hline
\noalign{\smallskip}
691 & 1994~${\rm Jan}$ ~10.48-11.69 & 19.81 \\
\hline
\noalign{\smallskip}
692 & 1994~${\rm Jan}$ ~11.82 - 12.77 & 15.87 \\
  \end{array} 
   \]
\end{table}

\begin{table}
\caption{Detected flux of GRS1009-45. Errors quoted are 1$\sigma$ errors.}
\label{Table2}
\[
    \begin{array}{cccc}
   \hline
\noalign{\smallskip}
${\rm Obs.~Session}$ & ${\rm Energy~band}$ & ${\rm Fluxes~(mCrab)}$ & ${\rm Errors}$ \\
\hline 
\noalign{\smallskip}
691 & 40-75 keV & 57 & 14.7 \\
\hline
\noalign{\smallskip}
691 & 75-150 keV & 39.4 & 20.4 \\
\hline
\noalign{\smallskip}
692 & 40-75 keV & 63.9 & 15.8 \\
\hline
\noalign{\smallskip}
692 & 75-150 keV & 97.8 & 22.7 \\
\hline
\noalign{\smallskip}
  \end{array} 
   \]
\end{table}

\begin{table}

\caption{Results of the spectral analysis for GRS1009-45. 1$\sigma $ errors
are quoted. The flux units are 10$^{-5}$ ph cm$^{-2}$s$^{-1}$ keV$^{-1}$.}
\label{Table3}
\[
    \begin{array}{ccc}
   \hline
\noalign{\smallskip}
\rm Power  & \rm F_{100~keV} & 3.51 \pm 0.5 \\
\noalign{\smallskip}
\rm Law   & \alpha & 1.9 \pm 0.4 \\
\noalign{\smallskip}
      & \chi^2_{\nu} & 0.78 (18~d.o.f.) \\
\hline
  \end{array} 
   \]
\end{table}

\noindent Figure 1 shows the source image in the 40-75 keV energy band.
GRS 1009-45 appears at a 6$\sigma$ confidence level. To estimate
the best source position, we selected data in the 60-120 keV energy
band. This choice was made in order to find a compromise between the higher
statistics of the low energy band and the narrower instrument PSF at
higher energies. A least square fit taking into account the instrumental PSF
gives a source position at 10h 11m 26.4s($\pm$ 4'), $ -$ 44 $^{\circ}$
49' 20"($\pm$ 4'), in complete agreement with the optical counterpart
position (Della Valle et al., 1993, 1997).

\noindent Table 2 shows the flux intensities of the source during the two
observing sessions. There is a considerable rise in the hard band flux during 
the second observation session, the flux rising from $\sim$ 40 mCrab to $\sim$
100 mCrab. However, the flux in the soft band is constant within the errors,
and the hardness ratios of the two observing sessions are respectively
HR(691) = 0.69 $\pm$ 0.4 and HR(692) = 1.53 $\pm$ 0.5. So there is no 
conclusive spectral change of the source between the two observations.
Given the relative faintness of the source, we consider this difference
not significant and the source flux being essentially constant during the
whole observation.

\noindent We thus summed the spectra obtained during the two observations.
We have performed spectral fit to the data using a power law model,
the results are presented in Table 3. The average spectrum is fitted
with a power law with spectral index $\alpha= -$ 1.9$\pm$0.4 and an integrated
energy flux of $\sim $ 6.9 ($\pm$ 1.7) $\times $ 10$^{-10}$ ergs cm$^{-2}$ 
s$^{-1}$ in the 40-150 keV energy band. The resulting spectrum is shown in Figure 2 along with the best fit, the parameters and errors of the fit are listed
in Table 3

\noindent The distance to the source has been estimated with the
optical observations by Della Valle et al. (1997) as being
between 1.5 and 4.5 kpc. Vargas et al. 1997 on the basis of the
hard X-ray outburst luminosity estimated a distance of 2.6 kpc,
with a 100 $\%$ uncertainty, fully compatible with the previous one.
Both of these values are compatible with the hydrogen absorption column
density previously quoted. Adopting these distance estimates we obtain an
X-ray luminosity L$_X$(40-150 keV) $\sim$10$^{35}$-10$^{36}$ ergs$^{-1}$.

\noindent This hard X-ray luminosity is similar to
the Nova Muscae one detected by SIGMA about 150 days after the outburst,
L$_x$(40-150 keV)$\sim$ 10$^{35}$ erg s$^{-1}$. (Goldwurm et al., 1993,
Sunyaev et al., 1992).

\begin{figure}

\centerline{\psfig{figure=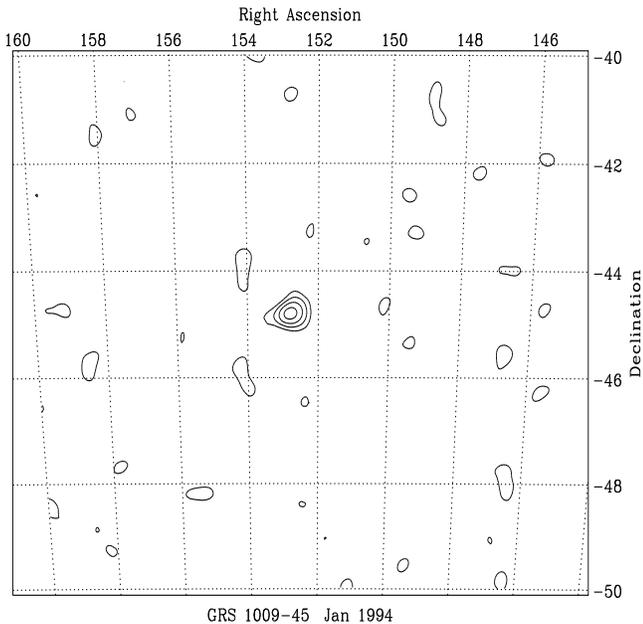,height=85mm,width=85mm,angle=90}}
\caption{Contour image of the sky region observed by SIGMA. Confidence 
levels start from 2$\sigma$ with 1$\sigma$ step.}

\label{Figure2}
\end{figure}




\begin{figure}

\centerline{\psfig{figure=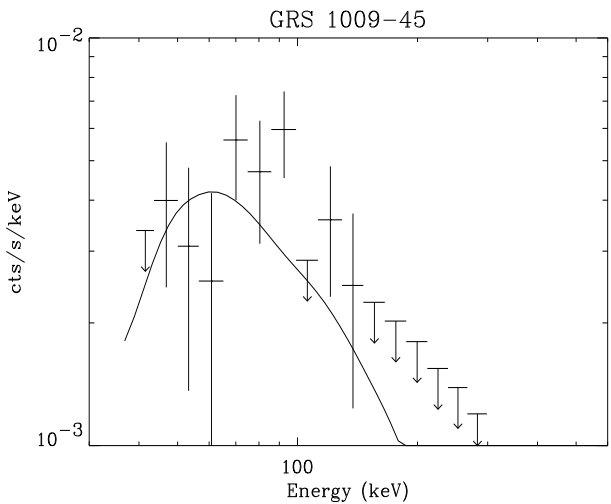,height=70mm,width=90mm}}
\caption{X-ray spectrum of GRS 1009-45 as detected by SIGMA.}

\label{Figure2}
\end{figure}

\section {Discussion}

\noindent X-ray Nova Velorum 1993 is a soft X-ray transient 
which displayed during its outburst a fair number of characteristics in
common with well known black hole Candidates. TTM and ASCA observations
shorly after the primary outburst showed that it displayed 
ultrasoft
X-ray emission along with a hard power law spectrum
extending up to $\sim$ 500 keV. This spectrum is thought to be a good
black hole indicator even if it is not considered to be conclusive 
(Tanaka $\&$ Shibazaki, 1996). Our observation showed that
the source emitted hard X-rays at least up to $\sim$ 150 keV
more than $\sim$ 100 days after the outburst.

\noindent The 2-200 keV outburst spectrum of GRS 1009-45 was
similar to the ones of GS2000+25 (Nova Vul 88, Tsunemi et al.,
1988) and GS/GRS1124-68 (Nova Mus 1991, Goldwurm et al., 1992,
Sunyaev et al. 1992, Ebisawa et al., 1994), both dynamically proven
black hole candidates (Remillard et al., 1992, Casares et al., 1995).
The primary outburst
was similar to the Nova Muscae one. However two later maxima in the
hard X-ray band occurred at $\sim$ 30 and $\sim$ 85 days after 
the primary maximum. Both occurred sooner than the secondary
maximum observed in GS1124-68 (Paciesas et al, 1995).
No type I X-ray burst, the best observational signature
of a neutron star in an X-ray transient, was detected.

\noindent Indeed a similar X-ray outburst light curve
characterized by a fast rise (few days) followed by a relatively
smooth exponential decay with a similar time constant (around
30-100 days) and at least one secondary outburst has been
observed in other X-ray novae both neutron stars and black
holes. In a recent review (Tanaka \& Shibazaki, 1996) four well known 
objects having a similar light curve are quoted. They are the above
mentioned GS2000+251 and GRS1124-68, A0620-00 and GROJ0422+32 and are
all strong black hole candidates. This classification is based on soft
(2-10 keV) monitoring for three of these objects, the exception being
GROJ0422+32, while for Nova Velorum 1993 only hard X-ray
monitoring has been performed. So while Nova Velorum cannot be
firmly associated at this group, it is worth mentioning
its similarity to these objects.

\noindent With this observation we monitored Nova Velorum 1993 during
the final phases of its outburst. Its luminosity was similar to the Nova
Muscae one as detected with SIGMA about 100 days after its primary outburst.
The hard X-ray spectral index that we detected, $\alpha$ =$-$ 1.9 $\pm$ 0.4,
is harder than the spectral index measured during the primary outburst
($\alpha=-$2.5$\pm$0.03 ) and it agrees with the spectral index
measured by BATSE after the December maximum.

\noindent We remark that a similar phenomenon possibly appeared also in
Nova Muscae 1991 for which Paciesas et al. (1995) and Ebisawa et al., (1994)
reported a harder X-ray spectrum near the end of the outburst. The same
phenomenon was detected by the SIGMA and ART-P instruments onboard the
GRANAT observatory (Gil'fanov et al., 1993b).


\noindent We conclude that the hard X-ray emission
of Nova Velorum 1993 looks similar to the ones of well established
black hole candidates, especially Nova Muscae 1991.

\noindent In summary, we have shown with the result of
SIGMA observations that GRS1009-45 was active over 100 keV
about 120 days after the primary outburst. Its hard X-ray
luminosity was comparable to the one of Nova Muscae $\sim$ 150 days after
the outburst. Its hard X-ray spectrum was
on average harder than the one detected during the primary outburst.

\noindent If we put our results together with all the other X-ray 
characteristics displayed by this source during its active phase,
mainly the outburst ultrasoft spectrum and the 20-100 keV light curve,
we believe that our observations strengthen the case for GRS 1009-45 to
be a black hole X-ray Nova.

\end{document}